# Evolution of low mass population III stars from the pre-main sequence to the white dwarf cooling track

T. M. Lawlor[1] and J. MacDonald[2]

[1] *Penn State University – Brandywine, Department of Physics, Media, PA 19063, USA*

[2] *University of Delaware, Department of Physics and Astronomy, Newark, DE 19716, USA*

**ABSTRACT**

Radiation feedback from massive population III stars may have given rise to low mass star formation from primordial or nearly primordial material. If early universe low mass stars did form, some should remain locally as white dwarfs, sub-giants, or main sequence stars. In this paper, we present model calculations for the evolution of single 0.8 $M_\odot$ – 3.0 $M_\odot$ stars with primordial metallicity from pre-main sequence to the white dwarf cooling track, and calculations for the evolution of single 4.0 $M_\odot$ – 7.0 $M_\odot$ stars which conclude in the giant phase. One goal of this work is to identify potential observable markers for potential observed progenitors of first or nearly first stars. We uncover a number of seemingly peculiar evolutionary differences between that of pop III low mass stars compared with younger higher $Z$ stars, as well as compared to other primordial evolution models. We also present an initial-final mass relationship and identify the minimum mass of a single white dwarf that could have had a population III progenitor**.**

## 1 INTRODUCTION

Early simulations of the formation of Population III stars predicted an Initial Mass Function (IMF) that is biased to high masses, $\gtrsim$ 100 $M_\odot$ (Bromm, Coppi & Larson 1999, 2002; Nakamura & Umemura 2001; Abel, Bryan & Norman 2000, 2002). Due to their short lifetimes, none of these stars will have survived to today, except as black hole remnants.

Recently, it has become clear that low mass early universe stars did form, as is evidenced in part by observations of low mass, very low metallicity stars, such as SDSS J102915+172927 (Caffau et al. 2011, 2012; Bonifacio, P. et al., 2018), HE0107-5240 (Christlieb et al. 2002), HD 140283 (Bond et al. 2013), J0023+0307 (Aguado et al. 2018; Frebel et al. 2019), SMSS J160540 (Nordlander et al. 2019),

SMSS J031300 (Keller et al. 2014), and others (Gallagher et al. 2010; Placco et al. 2013; Hansen et al. 2015; Norris 2018).

Stacy & Bromm (2014) showed that proto-stars from mini-haloes that form at redshift $z = 15$ accrete at much lower rates than in mini-haloes that form earlier at $z = 20 - 30$. The stars then end up having masses between < 1 $M_\odot$ and 5 $M_\odot$. Population III stars at the bottom of this mass range could survive to today evolving on the main sequence or on the Red Giant Branch (RGB). Depending on the amount of mass loss, the more massive of these low-mass stars could produce detectable white dwarf stars.

By assuming that protostars gain mass due to spherical Bondi–Hoyle accretion from the ambient medium, Dutta et al. (2020) find that Population III protostars with mass less than ~ 0.8 $M_\odot$ can be ejected from the star forming region and survive to the present day on the main sequence. From their simulations of star formation in primordial minihaloes of mass $10^5 - 10^6$ $M_\odot$, Latif, Whalen & Khochfar (2022) find that Pop. III stars have characteristic masses of $1 - 10$ $M_\odot$, and that ~70% of the stars are ejected from the minihaloes.

There is a more modest body of work dealing with observational signatures of pop III stars (Cayrel 1996; Schaerer & Pello 2002; Scannapieco, Schneider & Ferrara, 2003; Bromm & Larson 2004; Vangioni et al. 2011; Olave et al. 2019). Many of these focus on the effects of a top-heavy IMF, and predict expectations for reionization, background gravity waves, signatures in the 21 cm background and gamma ray bursts for example. Vangioni et al. explored the effects of cosmic chemical evolution when considering intermediate mass population III stars, similar to those in this paper. Olave et al. summarizes some observations and categories of metal poor stars, such as CEMPS and CEMP-no stars, which they (and others) suggest may be direct descendants of pop III stars. Their conclusions based on Europium detections in metal poor stars favor an IMF that is both intermediate (8-10 $M_\odot$, though they refer to it as low mass) and high mass (20-25 $M_\odot$). In their analysis, the lower limit in mass is not well constrained.

There is a growing body of work in modeling the evolution of low and intermediate mass Population III stars, which we briefly review here. Ezer & Cameron (1971) presented results for models in the mass range $5 - 200$ $M_\odot$. They found that because only the pp-chains generate energy their models had higher central temperatures than models with population I or II abundances. They also discovered that the $^{12}$C catalyst for the CNO-cycles was produced by the triple alpha process. Ezer (1972) found that the CNO-cycle turns on near the end of core hydrogen burning in a 3 $M_\odot$ model. Castellani & Paolicchi (1975) explored how the properties of main sequence Pop. III stars depended on the helium abundance. D'Antona & Mazzitelli (1982) followed the evolution of 0.9 and 1.0 $M_\odot$ models to the helium flash. They found that production of $^{12}$C by the 3α process led to a CNO-cycle driven hydrogen flash near the end of hydrogen burning. Similar results were found by Guenther & Demarque (1983). Analytic work by Fujimoto et al. (1984) predicted that there should be thresholds in core mass and CNO-abundance for helium pulses to occur. In a study of the evolution of a model of a 5 $M_\odot$ star, Chieffi & Tornambé (1984) found that the Thermal Pulsing Asymptotic Giant Branch (TP-AGB) phase did not occur, in agreement with the predictions of Fujimoto et al. (1984). Fujimoto, Iben & Hollowell (1990, also see Hollowell, Iben & Fujimoto 1990) found that in a 1 $M_\odot$ stellar model convection driven by the helium core flash penetrates into the H-rich envelope. Cassisi & Castellani (1993) evolved models with masses in the range 0.7 – 15 $M_\odot$ for heavy element abundances, $Z = 10^{-10}$, $10^{-6}$ and $10^{-4}$. They found that for their $Z = 10^{-10}$ models in which breathing pulses during core helium burning were suppressed the limiting mass ($M^{up}$) between development of a degenerate CO core or mildly degenerate, off center carbon burning, was 7.25 $M_\odot$ with an uncertainty of about 0.25 $M_\odot$. Cassisi et al. (1996) built on the work of Cassisi & Castellani (1993) with a focus on whether very-low metallicity models can produce currently observable RR Lyrae variables. Importantly, they found in their 0.8 $M_\odot$, $Z = 10^{-10}$ model a proton mixing event occurred during a strong helium shell flash that began under moderately degenerate conditions. Fujimoto, Ikeda & Iben (2000) studied the evolution of a 0.8 $M_\odot$ model in detail for $Z = 0$, [Fe/H] = -4, and [Fe/H] = -2. As did Fujimoto et al. (1990), they found that proton mixing occurred during the He core flash for their $Z = 0$ model. However, Weiss et al. (2000) did not find such mixing in their study of pop. III models of masses between 0.8 and 1.2 $M_\odot$. Schlattl et al. (2001) did find helium core flash mixing in models of 0.81, 0.82, and 1 $M_\odot$. In addition, Schlattl et al. followed the evolution of their 0.82 $M_\odot$ model with Reimers mass loss to the white dwarf stage. Schlattl et al. (2002) expanded on this work to further explore how the proton-mixing phenomenon depends on initial metallicity and treatment of convection. Chieffi et al. (2001) in models of 4, 5, 6, 7, and 8 $M_\odot$, found that their 8 $M_\odot$ model ignited carbon off-center in the early Asymptotic Giant Branch (AGB) phase, and contrary to the findings of Chieffi & Tornambé (1984), their 5 $M_\odot$ model did experience a TP-AGB phase. Marigo et al. (2001) evolved stellar models with masses in the range 0.7 – 100 $M_\odot$ without mass loss, up to the TP-AGB or C ignition. They found C-ignition occurred in models of mass 7 $M_\odot$ or higher. Siess, Livio & Lattanzio (2002) included convective core overshoot in their study of stellar models in the 0.8 – 20 $M_\odot$ mass range. They did not find proton mixing during the He core flash in their 1 $M_\odot$ model. However, Picardi et al. (2004) and Weiss et al. (2004) found the He-FDDM (He-flash driven deep mixing) phenomenon for $Z = 0$ models of mass M = 0.8 and 0.82 $M_\odot$ respectively. Suda, Fujimoto & Itoh (2007) found proton mixing during the He-core flash for M ≤ 1.1 $M_\odot$ but not in their M = 1.2 $M_\odot$ model in which the flash started at the stellar center. Campbell & Lattanzio (2008) studied Pop. III stars in the mass range 0.85 ≤ M ≤ 3 $M_\odot$ and included mass loss for the first time in order to determine nucleosynthetic yields for comparison with abundance determinations for extremely metal poor stars in the Galactic Halo. They introduced the terminology 'Dual Core Flash' (DCF) for the proton-mixing phenomenon when it occurs during the helium core flash and 'Dual Shell Flash' (DSF) for when it occurs during a helium shell flash (Cassisi et al. 1996; Chieffi et al. 2001; Siess et al. 2002). With the exception of a single model of Schlattl et al. (2001), in none of these studies of population III stars are the models evolved to the white dwarf stage.

Some low mass early universe stars should remain locally as white dwarfs or sub-giants, or for the lowest masses, MS red dwarfs. Iben & MacDonald (1986) showed that progenitors with lower CNO abundances depart the AGB with more massive H envelopes, and that this can lead to white dwarfs with thicker H envelopes. This finding was confirmed by Miller Bertolami, Althaus & García-Berro (2013), who found that their low metallicity, $Z = 0.001$, models arrived on the white dwarf (WD) cooling track with thicker hydrogen envelopes than their higher $Z$ counterparts. Following up on this, Althaus et al. (2015) modeled evolution for stars with metallicity as low as $Z = 3 \times 10^{-5}$. They found, due to thicker hydrogen envelopes, that their lowest metallicity models sustained stable hydrogen fusion well into the WD cooling track up to when $\log (L/L_\odot) = -3.2$. They concluded that this would have a significant impact in increasing the cooling time for low metallicity white dwarfs. In an earlier paper, Gil-Pons, Gutiérrez & García-Berro (2007) investigated the effects of convective overshoot on the evolution of models of primordial stars of mass between 5 and 10 $M_\odot$. They determined that including convective overshoot reduced the maximum initial mass at which CO cores and ONe cores develop, and so found that less massive ZAMS progenitors may result in a ONe WD or exploding directly as a Type II SN, rather than a CO WD.

We give a brief description of our evolutionary code in section 2, and present a grid of primordial progenitor evolution models in section 3. We follow our model evolution from ZAMS through the WD cooling track where possible, give our IFMR in section 4, along with the structure and compositions of models when the reach the WD track. We also identify the lowest mass main sequence star that reaches the WD cooling track within the time since star formation began. Finally, in section 5, we provide a summary and discussion.

## 2 STELLAR EVOLUTION CODE AND INITIAL MODELS

Our evolution models are calculated using the evolution code recently renamed DEuCES (DElaware Code for Evolution of Stars). Earlier versions of the DEuCES code are described in Lawlor & MacDonald (2003, 2006), and Lawlor et al. (2008, 2015). Some relevant aspects of the code are described in the next subsections. The mass range of the models explored here is from 0.80 to 7.0 $M_\odot$. Models less than or equal to 3.0 $M_\odot$ are evolved from the pre-main sequence to an endpoint on the white dwarf cooling track at age equal to the age of the Universe, which we take to be $1.37 \times 10^{10}$ yr. By interpolating in our set of models, we estimate that a model of mass 0.81 $M_\odot$ would be the lowest mass model to reach the white dwarf cooling track in time less than the first appearance of stars, approximately 200 million years after the big bang (Bowman et al. 2018). For population III composition, we adopt a metallicity of $Z = 10^{-14}$. Previous studies such as Siess et al. (2002) and Chieffi et al. (2001) adopted $Z = 0.0$, while Coc, Uzan & Vangioni (2014) took primordial composition to be $Z = 10^{-13}$ to $10^{-12}$ based on Big Bang nucleosynthesis calculations. Our $Z$ value is intermediate between the two. All population III models presented here begin with an initial hydrogen abundance of $X = 0.751$ and helium abundance $Y = 0.249$, based on big bang cosmology calculations (Casagrande et al. 2007).

### 2.1 Mass loss rates

For models with effective temperature $T_{eff} < 10,000$ K, we use the Reimers (1977) mass loss rate

$$\dot{M} = -4 \times 10^{-13} \eta \frac{L}{L_\odot} \frac{R}{R_\odot} \frac{M_\odot}{M} \ M_\odot \ \mathrm{yr}^{-1}, \tag{1}$$

with the Reimers' mass loss parameter set to $\eta = 0.378$. This particular value was chosen based on the mean value of $\eta$ needed to produce observed globular cluster horizontal branch (HB) morphologies (Jimenez et al. 1996). A number of the globular clusters analyzed by Jimenez et al. have since been found to have multiple populations (e.g. Gratton, Carretta & Bragaglia, 2012, and references therein). Thus, mass loss is likely to be only one of the physical processes responsible for HB morphology. In particular, from the location of the red giant bump in the cluster luminosity function, there is evidence that each population differs in their He abundance. The impact of variations in helium abundance on stellar models, including the HB morphology, has been recently reviewed by Cassisi & Salaris (2020). However, there is as yet no clear analysis of the interplay between assumed theoretical mass loss rates, such as the Reimers' law, and variations in the helium abundance on HB morphology. Such analysis would require investigation of the role played by the choice of mixing length parameter because of its major impact on the radius of the RGB stellar models (as shown by Jimenez et al. 1996), and hence on theoretical mass loss rate. The effects of different choices for the Reimers' mass loss parameter are touched on in section 3.

In addition to the Reimers mass loss rate, we use a mass loss rate for stars with dusty envelopes found by fitting to the dust mass loss rates derived by Groenewegen & Sloan (2018) for local group pulsating giant stars. We relate these mass-loss rates to the pulsation period in days, $P$, by

$$\dot{M} = -10^\alpha P^\beta \ M_\odot \ \mathrm{yr}^{-1}, \tag{2}$$

where $\alpha$ and $\beta$ are constants. We find the best fit parameters are $\alpha = -13.28$, $\beta = 2.34$ for M stars, and $\alpha = -17.06$, $\beta = 3.88$ for C stars. The C star mass loss rate is used only if the condensable carbon mass fraction exceeds a threshold value of 0.0019, which is based on the wind models of Mattsson, Wahlin & Hofner (2010). To determine the pulsation period from stellar parameters, we assume that AGB variables pulsate in the fundamental mode with period (Ostlie & Cox 1986)

$$P = 1.2 \times 10^{-2} \left(\frac{M}{M_\odot}\right)^{-0.73} \left(\frac{R}{R_\odot}\right)^{1.86}. \quad (3)$$

We use the dusty envelope mass loss rates when our models have properties similar to those of stars that exhibit dusty mass loss, e.g. pulsating stars on the TP-AGB. This includes phases in the evolution of our models which experience a DCF and do not have an AGB phase of evolution.

For evolutionary phases with $T_{eff} > 10^4$ K, the mass loss rate transitions to the OB star prescription of Vink, de Koter & Lamers (1999, 2000, 2001). We note that this prescription might overestimate the mass loss rates because of the absence of Fe in population III stars.

## 2.2 Convection

To determine the convective energy flux, we use the Schwarzschild criterion for convective onset,

$$\nabla > \nabla_{ad}, \quad (4)$$

and standard mixing length theory (Böhm-Vitense 1958) with a modification to include radiative losses from convective elements when they are optically thin (Mihalas 1978). Convective mixing is modelled as a diffusion process (Eggleton 1972), with turbulent diffusivity (Iben & MacDonald 1995)

$$\sigma_{con} = \beta_{con} w_{con} l, \quad (5)$$

where $l$ is the mixing length, $w_{con}$ is the convective velocity determined from the mixing length theory, and $\beta_{con}$ is a free parameter. Loosely based on turbulent transport ideas, we set $\beta_{con}$ to 1/3. To treat semiconvection as defined by Eggleton (1972) with our staggered mesh, we found that the opacity that enters into evaluation of $\sigma_{con}$ at the edges of convection zones has to be evaluated on the convective side of the boundary. This procedure leads to the difference between radiative and adiabatic gradients smoothly approaching zero as the edge of a convection zone as approached from the inside. Mixing due to convective overshooting is not included.

## 2.3 Opacity

Radiative opacities are determined by interpolation in the tables of Iglesias & Rogers (1996) for temperatures greater than 25,000 K and Ferguson et al. (2005) for temperatures less than 16,000 K, with a smooth interpolation of the two sets of opacities between these temperature limits. To include the effects of opacity enhancement due to convective dredge-up of carbon, we have scaled the low temperature opacities by comparison with opacity tables from Lederer & Aringer (2009). Conductive opacities are from Cassisi et al. (2007).

## 2.4 Nuclear reaction network, reaction rates and neutrino loss rates

Composition equations for H, D, $^3$He, $^4$He, $^7$Li, $^7$Be, $^{12}$C, $^{13}$C, $^{14}$N, $^{16}$O, $^{20}$Ne, $^{24}$Mg, $^{28}$Si and $^{56}$Fe are solved simultaneously with the structure and adaptive mesh equations. The reaction network includes 31 nuclear reactions with rates taken mainly from Angulo et al. (1999). For the $^{14}$N(p,γ)$^{15}$O reaction, we use the rate from Imbriani et al. (2005) as modified by Halabi, El Eid, & Champagne, (2012). The 3α, $^{12}$C(α,γ)$^{16}$O and $^{14}$N(α,γ)$^{18}$F rates are those of Fynbo et al. (2005), Katsuma (2012), and Gorres et al. (2000), respectively. Electron screening is treated by using the lowest of the weak screening, intermediate screening and strong screening factors of Salpeter (1954), Graboske et al. (1973), and Itoh et al. (1979,1990), respectively. Neutrino energy loss rates are from Itoh et al. (1996).

## 2.5. Computational method

The DEuCES (Delaware Code for Evolution of Stars) code has its origins in the Eggleton code which has significant differences from most other stellar evolution codes in that: 1) The equations of stellar structure are solved simultaneously with the composition equations, 2) The number of mesh points is fixed with the mass zoning determined by using an adaptive mesh method, and 3) All of the star inside the photosphere is included in the calculation without use of a separate envelope integration. Our code differs from the Eggleton code in that time derivatives are determined using a second-order upwind finite difference method instead of the original first order method. We found this is necessary to correctly follow the evolution during helium shell flashes.

The advantages and disadvantages of the opposite approach of separating the solution of the structure equations from solution of the composition equations in the context of convective-penetration of the overlying hydrogen-rich layer during the helium core flash are described in detail by Schlattl et al. (2001) in their section 2.1.

By comparing evolutionary tracks calculated with 500, 1000 and 2000 mesh points, we determined that 1000 mesh points is sufficient to give consistent results, and this is the number of points used for all of the models described here.

## 3 EVOLUTION OF THE MODELS

We find that our models of mass less than 1 M$_\odot$ experience a helium core flash which leads to a DCF. Following the DCF, significant convective dredge up of CNO elements occurs which increases the opacity of the envelope material. As a consequence, the star expands and evolves to the red. Due to expansion of the envelope helium burning is temporarily quenched. These models experience an extended phase of steady shell hydrogen burning before helium burning resumes. At the reddest phase of the evolution, the model has surface properties similar to those of cool dusty AGB stars and because of the dredge-up of CNO elements following the DCF with the carbon abundance being larger than the oxygen abundance by a factor of about 3, dust grains are likely to form. Our dusty mass loss formula leads to significant mass loss during the hydrogen burning phase. For models of mass less than or equal to 1.0 $M_\odot$, the mass loss reduces the envelope mass to the extent that a helium shell burning phase does not occur. In the case of our 0.80 $M_\odot$ model, the mass lost is sufficiently large that helium burning does not rekindle after the hydrogen burning phase and the model becomes a helium white dwarf.

Because of uncertainties associated with mass loss, particularly during the 'dusty' phase of evolution predicted by our models, we also consider for purposes of comparison with earlier work the evolution of a 0.85 $M_\odot$ model that does not include dusty mass loss. Also, because recent asteroseismic analysis by Miglio et al (2012) indicates that setting $\eta \sim 0.30$ might be more appropriate, we have calculated the evolution of a 0.85 $M_\odot$ model that does not include dusty mass loss with $\eta = 0.30$.

### 3.1 Evolution of a 0.80 $M_\odot$ model

Although our 0.80 $M_\odot$ model does not reach the white dwarf cooling track in the age of the Universe, we present results for this mass for purposes of comparison with other work and the evolution of our models of higher mass. In figure 1, we show the model's evolutionary track in the Hertzsprung-Russell diagram.

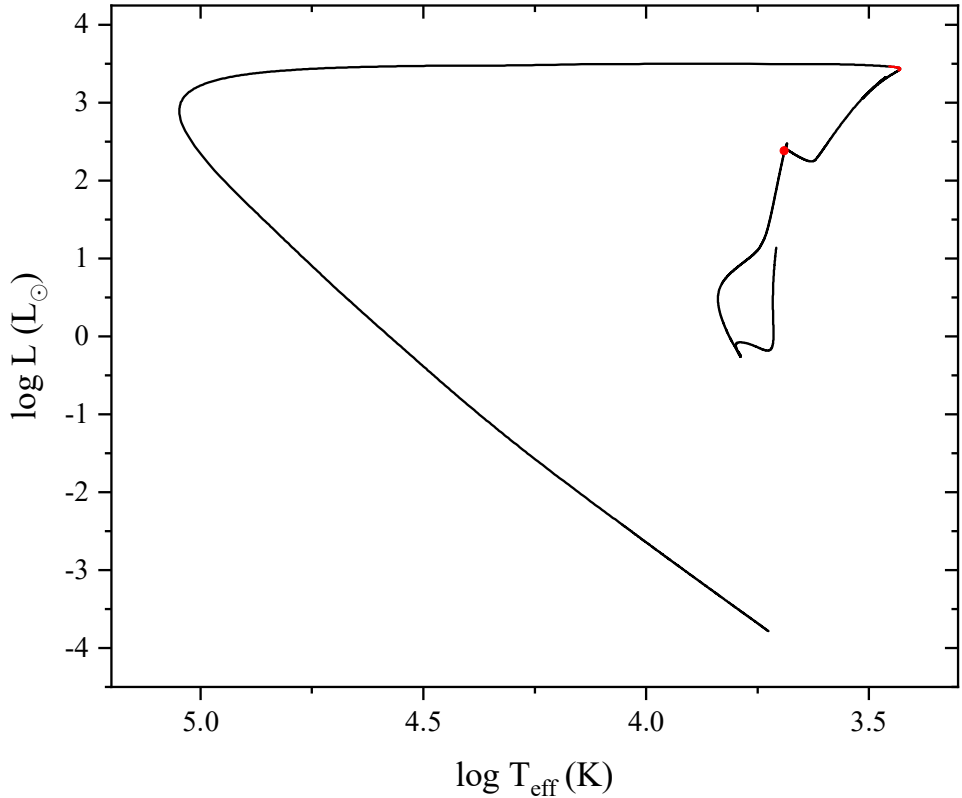

Figure 1. Evolution in the HRD of the 0.80 M$_\odot$ model. The red circle shows the point at which the core helium flash begins. The part of the evolutionary track in red shows where mass loss is mainly due to a dusty wind.

The helium core flash begins off-center at age 14.0 Gyr. Convection driven by the helium core flash penetrates into the H-rich envelope, starting shortly before the peak in helium burning energy generation rate. Inward mixing of protons to deeper, hotter layers leads to a rapid increase in the hydrogen burning luminosity, and the energy generation rate from hydrogen burning reaches a peak of $6.30\times10^{11}$ L$_\odot$, exceeding the peak helium burning energy generation rate of $2.65\times10^{10}$ L$_\odot$. Expansion and cooling of the envelope leads to a decrease in helium burning. Convective dredge up increases the surface abundances of the CNO elements to $Z_{CNO} \sim 0.014$, with nitrogen being the most abundant element due to CNO-processing of carbon produced by the 3α process. Because of the resulting increase in opacity the star increases in radius and evolves to the red, as clearly seen in figure 1. The surface hydrogen and helium abundances at this point are X = 0.48 and Y = 0.51. A phase of hydrogen-shell burning ensues, lasting ~2.5 Myr. At this point the model is evolving in a similar way to an RGB star. Substantial mass loss occurs during this hydrogen-shell burning phase, reducing the stellar mass from 0.763 M$_\odot$ to 0.456 M$_\odot$. Approximately 0.104 M$_\odot$ is lost during a phase near the tip of the RGB (shown in red in figure 1) in which dusty wind mass loss dominates. The final mass is low enough that the central temperature does not get high enough for helium burning to rekindle and the model ends as a He WD. At the point of departure from the RGB, mass loss and shell hydrogen burning have reduced the hydrogen envelope mass to ~0.06 $M_\odot$. With further mass loss and increasing effective temperature, the model enters a phase similar to that of a CSPN that lasts ~$3\times10^4$ yr.

We find that during the dredge up phase the surface Li mass fraction increases to ~$10^{-5}$, so that approximately $2\times10^{-6}$ $M_\odot$ of Li is returned to the interstellar medium during the hydrogen shell burning stage. This is more than the mass of Li locked up during formation of the star, and hence low mass Pop. III stars could be a Li source.

Picardi et al. (2004) have also computed the evolution of a 0.8 $M_\odot$ Pop III model. However, they did not include mass loss, and hence do not follow the evolution to the white dwarf stage. The evolution of their model is similar to what we find for the early evolution in which mass loss is negligible. Picardi et al. found that a DCF occurs and leads to dredge-up of CNO isotopes. However, because they use metal-free low-temperature opacities after dredge-up, Picardi et al. do not find the evolution to the red that occurs in our model. They found that $^7$Li is produced in large amounts during the DCF, reaching a surface abundance of log $\varepsilon$ = 5.87, which is comparable to the surface value found in our calculations.

### 3.2.1 Evolution of our 0.85, 0.90, 0.95 and 1.00 $M_\odot$ models

The initial phases of the evolution of our 0.85, 0.90, 0.95 and 1.00 $M_\odot$ models are qualitatively similar to those of the 0.80 $M_\odot$ model in that they all experience a DCF. However, the stellar mass at the end of the post He core flash hydrogen shell burning phase is high enough for He burning to rekindle. Due to significant mass loss after the DCF, none of these models retain an envelope that is sufficiently massive for there to be an AGB phase.

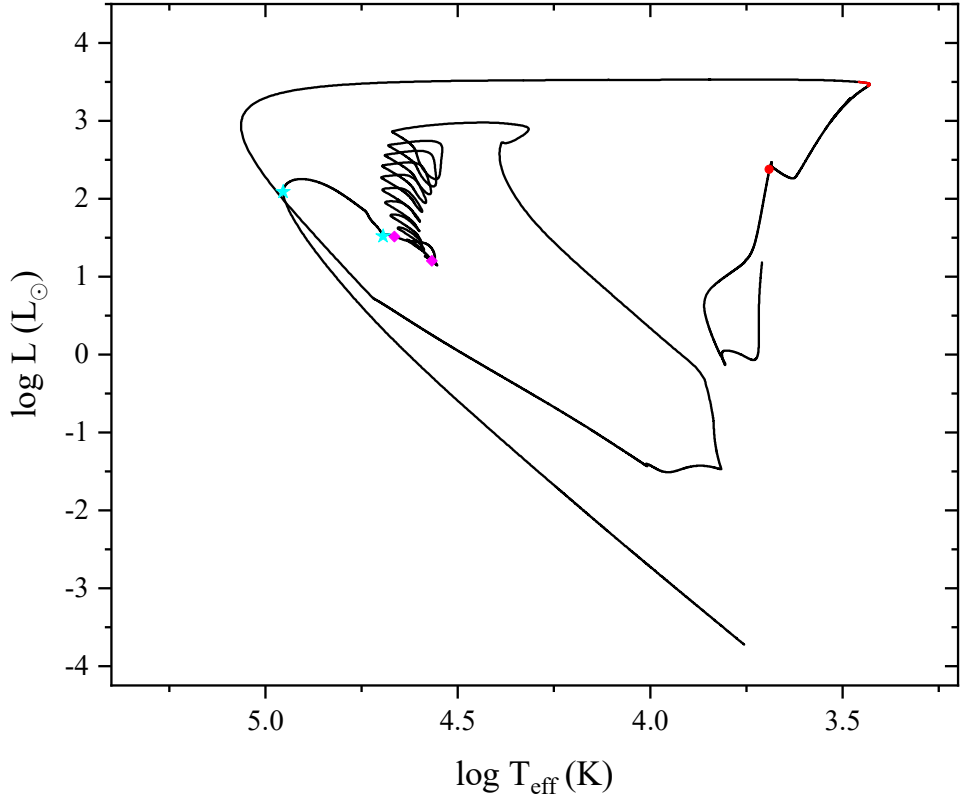

Figure 2. Evolution in the HRD of our 0.85 M$_\odot$ Pop III model. The red circle shows the point at which the core helium flash begins. The He core burning and He shell burning phases are delineated by magenta diamonds and light blue asterisks, respectively. The part of the evolutionary track in red shows where mass loss is mainly due to a dusty wind.

The HRD evolutionary track of our 0.85 M$_\odot$ model is shown in figure 2. When the model reaches the tip of the RGB mass loss has reduced the mass of the hydrogen envelope to 0.12 M$_\odot$ and convective dredge-up has increased the total CNO mass fraction to 0.0125. The total mass lost during the phase in which the dusty wind is the dominant (but not only) source of mass loss is 0.123 M$_\odot$. Mass loss and hydrogen burning episodes between the secondary helium core flashes further reduce the hydrogen envelope mass to $9\times10^{-3}$ M$_\odot$ at the point at which core helium burning begins. Due to wind mass loss carrying away much of the envelope, at this stage the surface composition is mainly helium with Y = 0.90, X = 0.056, and $Z_{CNO}$ = 0.043.

Because of the low hydrogen rich envelope mass ($M_{Henv} \sim 3\times10^{-4}$ M$_\odot$), the HRD loops due to the secondary core helium flashes occur at high effective temperature, $T_{eff} \sim$ 35,000 – 50,000 K. This behaviour is similar to that found in the 'hot flasher' scenario first introduced by Castellani & Castellani (1993), and later used for creating extreme horizontal branch (EHB) stars (D'Cruz et al. 1996) by increasing the standard Reimers mass loss rate by factor of order 2. First fully evolved by Cassisi et al. (2003), similar models used were used by Rodríguez-López et al. (2010) to analyze pulsations of sdO stars. In that paper, low mass sdO models were created by evolving models of initial mass 1.0 $M_\odot$ with the standard Reimers mass loss rate increased by a factor of 1.6 – 1.85. Here we see that a similar result is obtained for population III stars of initial mass ~0.85 $M_\odot$ in which there is an epoch of mass loss initiated by proton-mixing during the helium core flash at a rate which is on average 25% greater than the standard Reimers mass loss rate. The model has an EHB phase that lasts approximately 130 Myr, with the more luminous helium shell burning sdO phase lasting about 23 Myr.

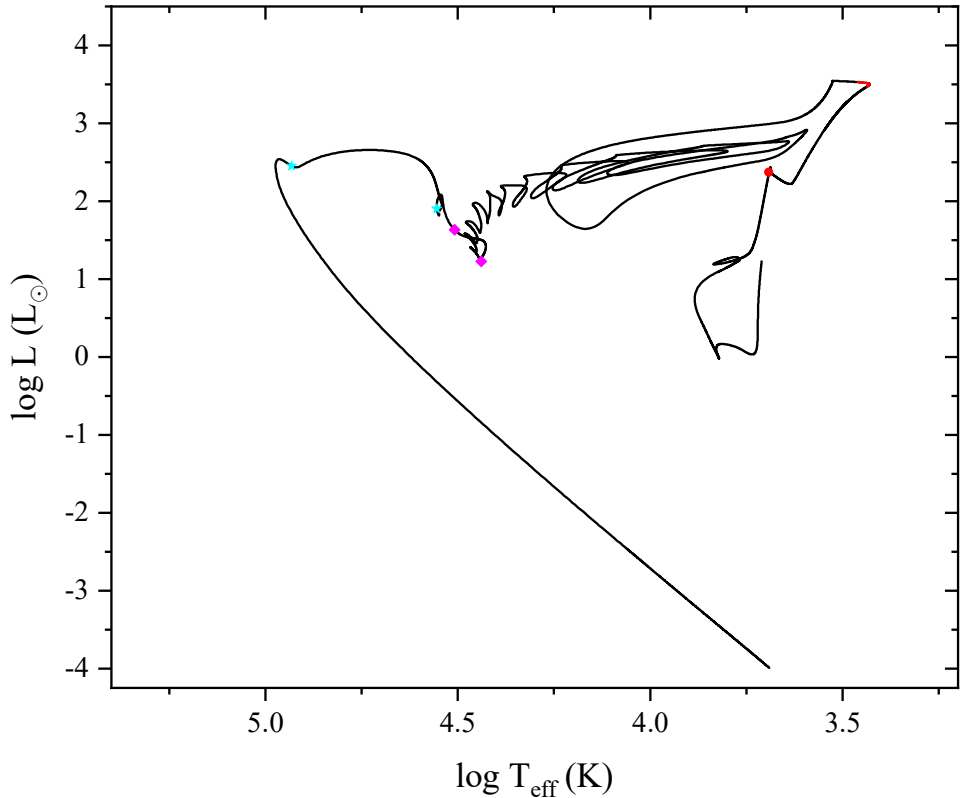

Figure 3. Evolution in the HRD of our 0.90 $M_\odot$ Pop III model. The red circle shows the point at which the core helium flash begins. The He core burning and He shell burning phases are delineated by magenta diamonds and light blue asterisks, respectively. The part of the evolutionary track in red shows where mass loss is mainly due to a dusty wind.

The HRD evolutionary tracks of our 0.90 and 0.95 $M_\odot$ models are shown in figures 3 and 4 respectively. Because of its larger mass, helium burning rekindles at a lower effective temperature in our 0.90 $M_\odot$ model than the 0.85 $M_\odot$ model. However, at the end of the secondary core flashes, the effective temperature has increased to ~27,000 K. The model then evolves through EHB and sdO phases before cooling as a WD.

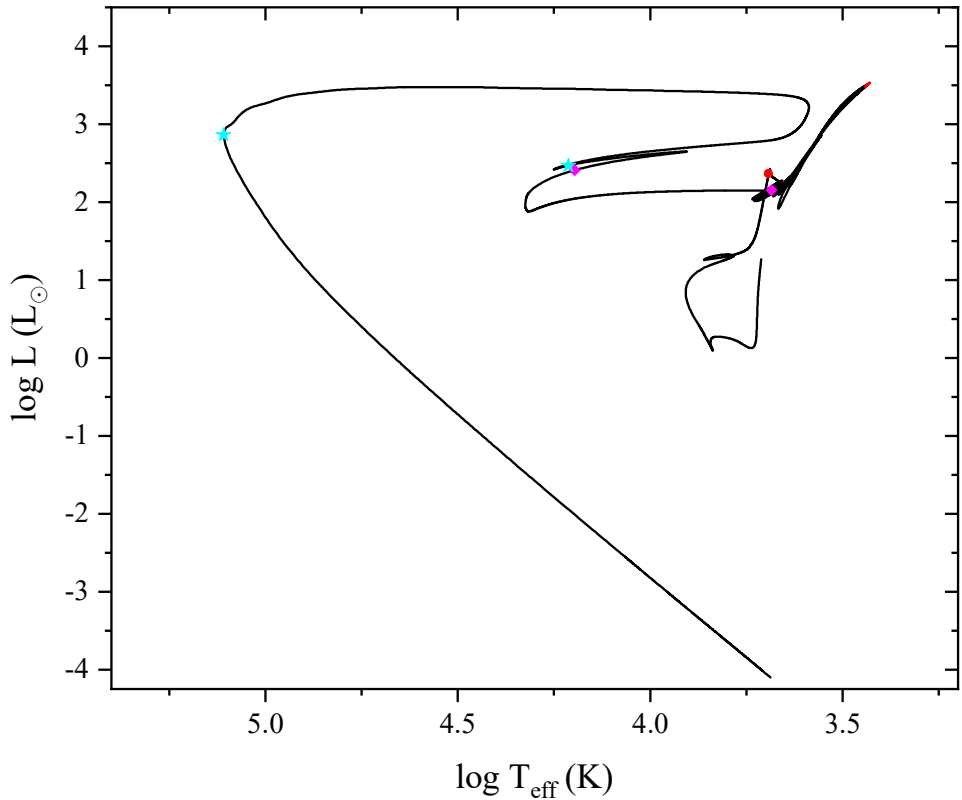

Figure 4. Evolution in the HRD of our 0.95 M$_\odot$ Pop III model. The red circle shows the point at which the core helium flash begins. The He core burning and He shell burning phases are delineated by magenta diamonds and light blue asterisks, respectively. The part of the evolutionary track in red shows where mass loss is mainly due to a dusty wind.

For our 0.95 M$_\odot$ model helium burning rekindles while still on the RGB. The secondary core flashes all occur at low $T_{eff}$, and at their end the model enters a HB phase. The surface He mass fraction in the HB phase, $Y = 0.44$, is high enough for an extended blue loop to occur (Sweigart & Gross 1976). During the sdO phase, the model evolves to the red but does not reach the AGB.

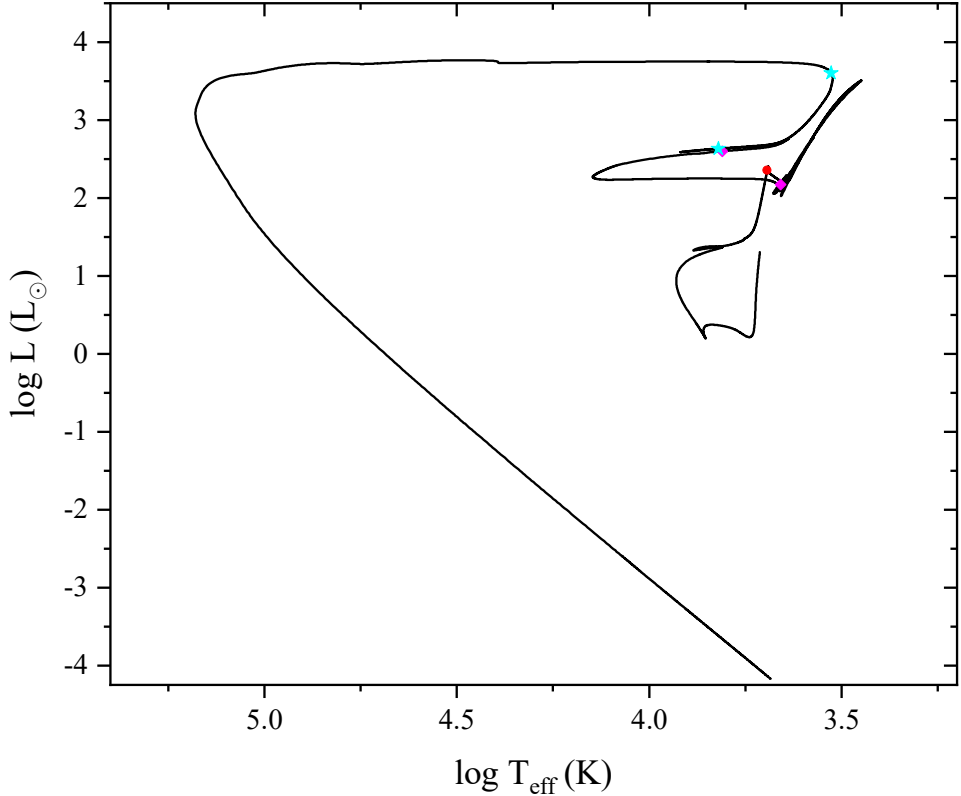

Figure 5. Evolution in the HRD of our 1.00 M$_\odot$ Pop III model. The red circle shows the point at which the core helium flash begins. The He core burning and He shell burning phases are delineated by magenta diamonds and light blue asterisks, respectively.

The evolutionary track in the HRD for our 1.00 M$_\odot$ is similar to that of the 0.95 M$_\odot$ model. However, at no time during the evolution does the dusty wind provide the majority of the mass loss. Very near the end of core hydrogen burning in our 0.90, 0.95 and 1.00 M$_\odot$ models, we find that sufficient $^{12}$C has been produced at the stellar center by the 3α process for the dominant nuclear energy generation process to switch from the pp-chains to the CNO-cycles. A convection zone develops and grows outwards causing a small amount of mixing in the tail of the hydrogen distribution. The resulting increase in nuclear energy generation is responsible for the loops at the end of the main sequence phase in figures 3, 4 and 5.

Table 1 summarizes some of the properties of the models described in this section. The first seven column headings are: 1) the initial mass, $M_i$, 2) the age at the He core flash, $\tau_{cf}$, 3) the peak He burning luminosity, $L_{Hemax}$, 4) the peak H burning luminosity in the DCF, $L_{Hmax}$, 5) the final WD mass, $M_f$, 6) Age at the beginning of the WD cooling phase, $\tau_{bwd}$, 7) the surface Z after the DCF, $Z_{DCF}$. Columns 8) to 11) are the total amounts of $^7$Li, $^{12}$C, $^{14}$N, and $^{16}$O returned to the ISM, respectively.

Table 1. Summary of some properties of models that experience a DCF.

| $M_i$ | $\tau_{cf}$ | $L_{Hemax}$ | $L_{Hmax}$ | $M_f$ | $\tau_{bwd}$ | $Z_{DCF}$ | $^7Li_{yield}$ | $^{12}C_{yield}$ | $^{14}N_{yield}$ | $^{16}O_{yield}$ |
| --- | --- | --- | --- | --- | --- | --- | --- | --- | --- | --- |
| ($M_\odot$) | (Gyr) | ($10^9 L_\odot$) | ($10^{11} L_\odot$) | ($M_\odot$) | (Gyr) | | ($M_\odot$) | ($M_\odot$) | ($M_\odot$) | ($M_\odot$) |
| 0.80 | 14.01 | 6.43 | 6.30 | 0.456 | 14.01 | 0.0144 | $1.9\times10^{-6}$ | $1.4\times10^{-3}$ | $2.2\times10^{-3}$ | $3.8\times10^{-4}$ |
| 0.85 | 11.28 | 6.19 | 6.03 | 0.471 | 11.44 | 0.0125 | $1.8\times10^{-6}$ | $1.3\times10^{-3}$ | $2.3\times10^{-3}$ | $3.6\times10^{-4}$ |
| 0.90 | 9.22 | 5.67 | 5.76 | 0.494 | 9.36 | 0.0112 | $7.2\times10^{-7}$ | $1.2\times10^{-3}$ | $2.3\times10^{-3}$ | $3.4\times10^{-4}$ |
| 0.95 | 7.63 | 5.16 | 5.42 | 0.590 | 7.72 | 0.0102 | $2.5\times10^{-7}$ | $9.4\times10^{-4}$ | $1.9\times10^{-3}$ | $2.7\times10^{-4}$ |
| 1.00 | 6.39 | 4.59 | 4.97 | 0.651 | 6.47 | 0.0093 | $9.5\times10^{-8}$ | $7.9\times10^{-4}$ | $1.7\times10^{-3}$ | $2.3\times10^{-4}$ |

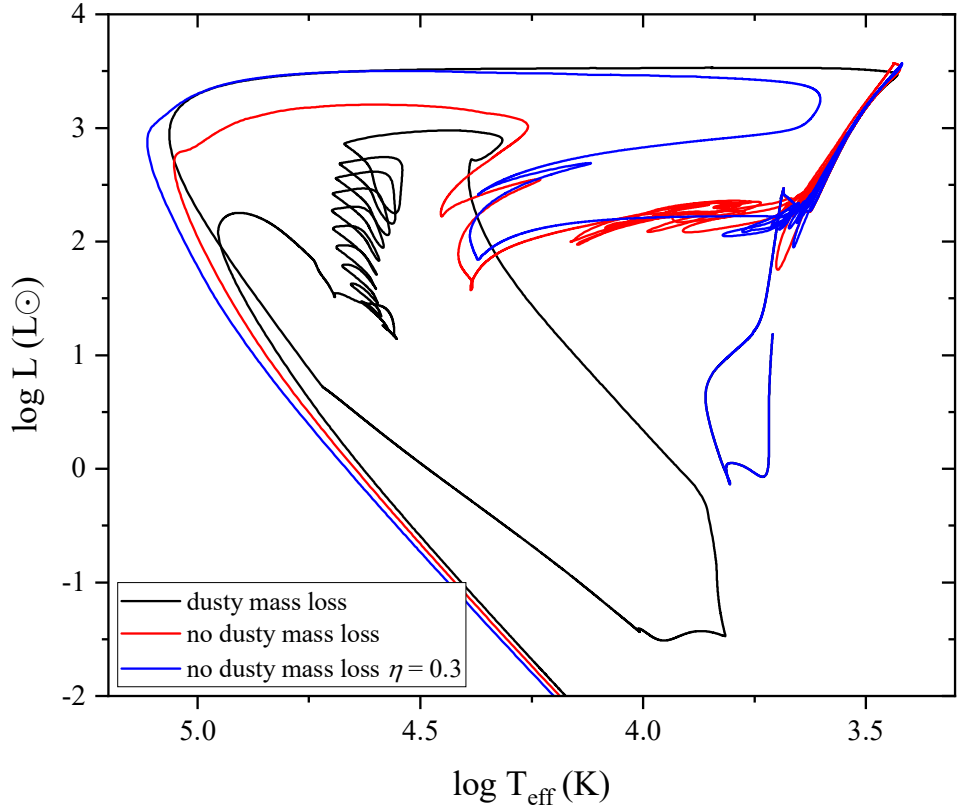

Figure 6. Evolution in the HRD of 0.85 $M_\odot$ Pop III models with and without dusty mass loss.

### 3.2.2 Evolution of models of mass 0.85 $M_\odot$ without dusty mass loss

To investigate the effects of different treatments of mass loss, we have evolved 0.85 $M_\odot$ models without dusty mass loss for Reimers parameter values $\eta = 0.378$ and $0.300$. Figure 6 compares the HRDs for the evolution of these models with that when dusty mass loss is included. The models without dusty mass loss have higher hydrogen rich envelope masses after the DCF, and hence the HRD loops due to the secondary core helium flashes occur at lower effective temperature. The reduction in mass loss rate leads to higher white dwarf mass. We find $M_f$ = 0.545 and 0.593 M$_\odot$ for $\eta$ = 0.378 and 0.300, respectively. To see when significant mass loss occurs, we give in Table 2 the stellar mass at a few evolutionary points.

Table 2. Stellar mass at select evolutionary points for models of initial mass 0.85 M$_\odot$ and different mass loss prescriptions.

| Mass loss prescription | Mass at He ignition (M$_\odot$) | Mass at end of DCF (M$_\odot$) | Mass at beginning of quiescent core He burning (M$_\odot$) | Mass at end of quiescent core He burning (M$_\odot$) |
| --- | --- | --- | --- | --- |
| Dusty, $\eta$ = 0.378 | 0.8171 | 0.4713 | 0.4712 | 0.4711 |
| No Dusty, $\eta$ = 0.378 | 0.8171 | 0.5499 | 0.5485 | 0.5484 |
| No Dusty, $\eta$ = 0.300 | 0.8240 | 0.6230 | 0.6211 | 0.6171 |

Approximately 0.03 M$_\odot$ is lost on the RGB before the onset of He burning. Most mass is lost during the DCF phase due the extended period of hydrogen burning. Also, Reimers' mass loss contributes the most to the total mass loss during this phase. After the core helium burning phase, the no dusty wind, $\eta$ = 0.300 model evolves to cool enough surface temperatures that Reimers' mass loss is sufficient to further reduce the mass by 0.024 M$_\odot$.

In table 3, we compare the same properties as given in Table 1 for the three cases of the evolution of the 0.85 M$_\odot$ models.

Table 3. As table 1 except just for the 0.85 M$_\odot$ models.

| Sequence | $\tau_{cf}$ (Gyr) | $L_{Hemax}$ ($10^9$ L$_\odot$) | $L_{Hmax}$ ($10^{11}$ L$_\odot$) | $M_f$ (M$_\odot$) | $\tau_{bwd}$ (Gyr) | $Z_{DCF}$ | $^7$Li$_{yield}$ (M$_\odot$) | $^{12}$C$_{yield}$ (M$_\odot$) | $^{14}$N$_{yield}$ (M$_\odot$) | $^{16}$O$_{yield}$ (M$_\odot$) |
| --- | --- | --- | --- | --- | --- | --- | --- | --- | --- | --- |
| Dusty, $\eta$ = 0.378 | 11.28 | 6.19 | 6.03 | 0.471 | 11.44 | 0.0125 | 1.8×10$^{-6}$ | 1.3×10$^{-3}$ | 2.3×10$^{-3}$ | 3.6×10$^{-4}$ |
| No Dusty, $\eta$ = 0.378 | 11.28 | 6.19 | 6.03 | 0.545 | 11.38 | 0.0125 | 1.4×10$^{-6}$ | 1.0×10$^{-3}$ | 1.8×10$^{-3}$ | 2.8×10$^{-4}$ |
| No Dusty, $\eta$ = 0.300 | 11.26 | 6.07 | 6.10 | 0.593 | 11.34 | 0.0124 | 1.1×10$^{-6}$ | 8.4×10$^{-4}$ | 1.5×10$^{-3}$ | 2.4×10$^{-4}$ |

### 3.3 Evolution of models of mass between 1.10 and 3.00 M$_\odot$

None of our models of mass less than 1.00 M$_\odot$ experience a TP-AGB phase. In this section, we describe the evolution of models of mass between 1.10 M$_\odot$ and 3.00 M$_\odot$ that do experience a TP-AGB phase. In each case, He burning begins at the center, under conditions of mild electron degeneracy or non-degeneracy and consequentially any core flash is weak and does not lead to a DCF. During the first helium shell flash, the flash-driven convection zone extends into the hydrogen layer leading to a DSF, following which CNO-nuclei are dredged to the surface. In a similar manner to the DCF models, the resultant increase in opacity leads to an increase in radius and a decrease in effective temperature.

To illustrate the typical evolutionary behaviour, we show in figure 7 the HRD track for our 1.3 $M_\odot$ model. The red circles show the core helium burning phase, and the magenta diamond shows where the first helium shell begins. Ingestion of protons into flash-driven convection zone occurs when $L_{He} \sim 4\times10^5$ $L_\odot$, leading to a peak $L_H \sim 2\times10^{10}$ $L_\odot$. Expansion of the envelope shuts off the He burning shell, and a quiescent H burning phase of duration $7\times10^5$ yr ensues. At the end of the H burning phase, He burning rekindles and reaches a peak $L_{He} \sim 3\times10^7$ $L_\odot$. Convective dredge up of CNO elements to the surface occurs a few hundred years after the peak in H burning. For the 1.3 $M_\odot$ model, the surface abundances become mainly $^{14}N$ with a total CNO mass fraction of $Z_{CNO} = 9\times10^{-4}$. Although this model experiences 29 additional thermal pulses, none causes further dredge up and the thermal pulses follow a similar path in the HRD until the envelope mass becomes small enough that blueward evolution occurs. However, in models of initial mass 1.5 $M_\odot$ and greater, further dredge up episodes occur and the TP-AGB moves redward. This behaviour can clearly be seen in the HRD for our 2.0 $M_\odot$ model shown in figure 8.

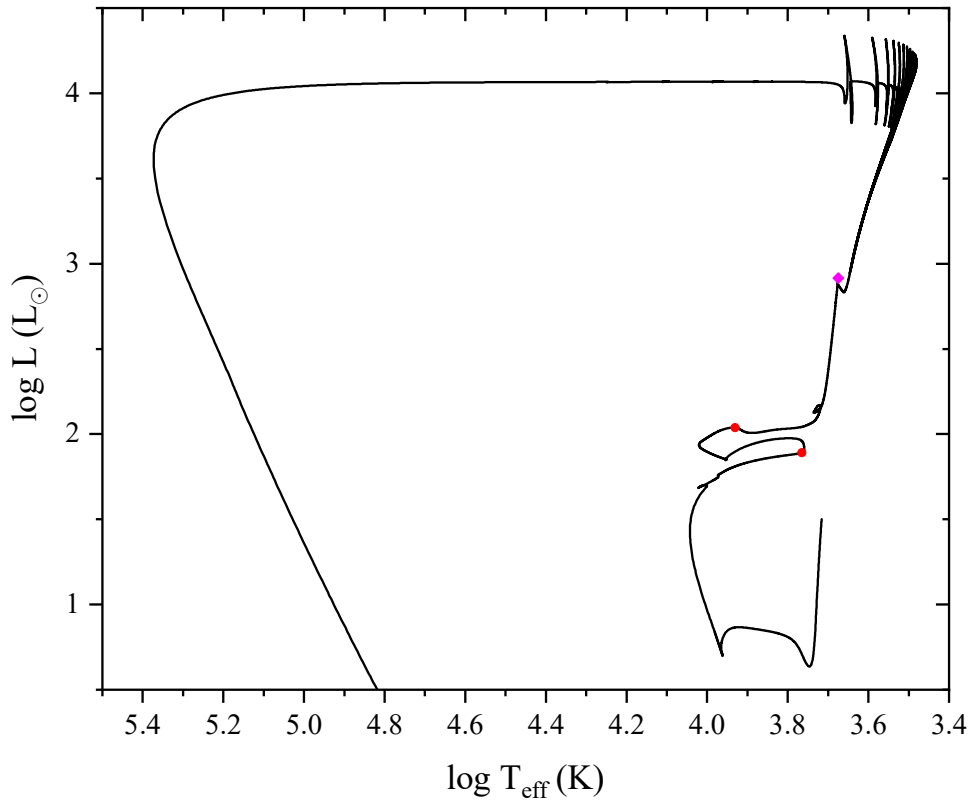

Figure 7. Evolution in the HRD of our 1.30 $M_\odot$ Pop III model. The red circles bracket the core helium burning phase, and the magenta diamond indicates the onset of the first helium shell flash, which becomes a DSF.

Also, we find that in models of initial mass 2.0 $M_\odot$ and greater, the dominant dredge up CNO isotope becomes $^{12}$C after the second flash. Near the end of the TP-AGB phases of our 2.5 and 3.0 $M_\odot$ models, the outer envelope becomes unstable with velocities exceeding the sound speed. As this occurs at the luminosity peak of a thermal pulse cycle, the instability is likely due to the Hydrogen Recombination Instability (Wagenhuber & Weiss 1994). To be able to continue the evolution to the white dwarf stage, we impose an external pressure of ~$10^2$ dyne cm$^{-2}$. To estimate the impact of adding the external pressure, we compare the mass of the hydrogen-exhausted core at the point at which instability occurs in the absence of the external pressure to the final white dwarf mass obtained by imposing the external pressure (the values given in table 3). The mass difference in each case is ~0.03 $M_\odot$, which we consider as an estimate of the uncertainty in the final white dwarf mass for these two models.

Our 3.0 $M_\odot$ pop III model experiences a DSF during which dredge up of CNO isotopes occurs. The major dredge up species in the first flash is $^{14}$N. Further significant dredge up of CNO isotopes occurs in subsequent helium shell flashes, with $^{12}$C being the dominant isotope. Table 4 summarizes some of the properties of the models described in this section. The first seven column headings are: 1) the initial mass, $M_i$, 2) the age at the first He shell flash, $\tau_{sf1}$, 3) the peak He burning luminosity just before DSF, $L_{Hemax}$, 4) the peak H burning luminosity in DSF, $L_{Hmax}$, 5) the final WD mass, $M_f$, 6) the age at the beginning of the WD cooling phase, $\tau_{bwd}$, and 7) the surface Z after the DSF, $Z_{DSF}$. Columns 8) to 11) are the total amounts of $^7$Li, $^{12}$C, $^{14}$N, and $^{16}$O returned to the ISM, respectively.

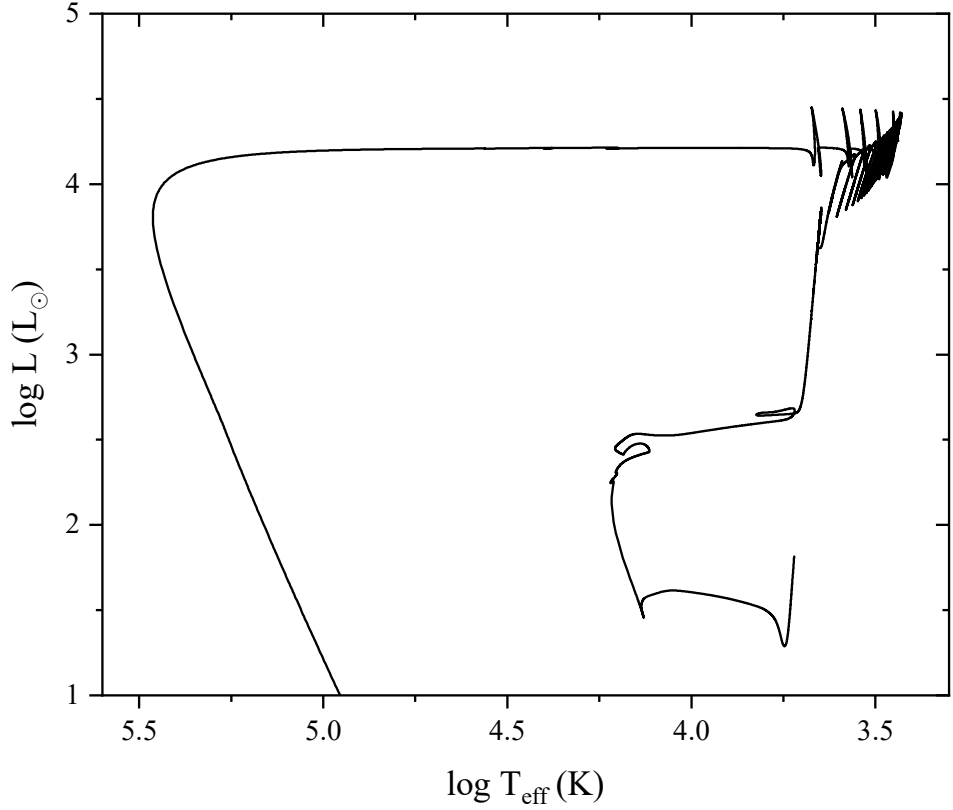

Figure 8. HRD evolutionary track of the 2.0 $M_\odot$ model.

Table 4. Summary of some properties of models that experience a DSF and are evolved to the white dwarf stage.

| $M_i$ | $\tau_{sfl}$ | $L_{Hemax}$ | $L_{Hmax}$ | $M_f$ | $\tau_{bwd}$ | $Z_{DSF}$ | $^7Li_{yield}$ | $^{12}C_{yield}$ | $^{14}N_{yield}$ | $^{16}O_{yield}$ |
|---|---|---|---|---|---|---|---|---|---|---|
| ($M_\odot$) | (Gyr) | ($10^6 L_\odot$) | ($10^{10} L_\odot$) | ($M_\odot$) | (Gyr) | | ($M_\odot$) | ($M_\odot$) | ($M_\odot$) | ($M_\odot$) |
| 1.10 | 4.746 | 1.57 | 2.29 | 0.648 | 4.753 | $8.1\times10^{-4}$ | $2.1\times10^{-9}$ | $4.7\times10^{-5}$ | $3.2\times10^{-4}$ | $1.6\times10^{-5}$ |
| 1.20 | 3.563 | 0.529 | 2.90 | 0.680 | 3.569 | $9.6\times10^{-4}$ | $2.4\times10^{-9}$ | $5.1\times10^{-5}$ | $4.0\times10^{-4}$ | $2.0\times10^{-5}$ |
| 1.30 | 2.751 | 0.371 | 2.68 | 0.715 | 2.755 | $8.9\times10^{-4}$ | $1.9\times10^{-9}$ | $5.4\times10^{-5}$ | $4.2\times10^{-4}$ | $1.9\times10^{-5}$ |
| 1.40 | 2.176 | 0.171 | 2.87 | 0.743 | 2.180 | $8.1\times10^{-4}$ | $1.4\times10^{-9}$ | $1.9\times10^{-4}$ | $4.4\times10^{-4}$ | $2.6\times10^{-5}$ |
| 1.50 | 1.755 | 0.022 | 1.49 | 0.771 | 1.758 | $7.6\times10^{-4}$ | $9.5\times10^{-10}$ | $3.5\times10^{-4}$ | $4.8\times10^{-3}$ | $3.3\times10^{-5}$ |
| 1.75 | 1.088 | 0.093 | 1.70 | 0.805 | 1.089 | $5.1\times10^{-4}$ | $3.3\times10^{-9}$ | $1.6\times10^{-3}$ | $3.1\times10^{-4}$ | $8.9\times10^{-5}$ |
| 2.00 | 0.7226 | 0.170 | 0.809 | 0.782 | 0.7233 | $1.8\times10^{-4}$ | $3.2\times10^{-9}$ | $2.6\times10^{-3}$ | $1.9\times10^{-4}$ | $9.8\times10^{-5}$ |
| 2.50 | 0.3806 | 0.0895 | 1.26 | 0.842 | 0.3813 | $1.4\times10^{-4}$ | $2.8\times10^{-9}$ | $5.9\times10^{-3}$ | $2.3\times10^{-4}$ | $2.2\times10^{-4}$ |
| 3.00 | 0.2341 | 4.88 | 0.129 | 0.885 | 0.2348 | $6.2\times10^{-5}$ | $4.6\times10^{-10}$ | $1.2\times10^{-2}$ | $1.1\times10^{-4}$ | $5.1\times10^{-4}$ |

**3.4 Evolution of models of mass greater than 3.00 $M_\odot$**

Here we consider the evolution of higher mass models for which the evolution tracks are incomplete and end before the white dwarf cooling track. Our goal in calculating these tracks is to determine mass limits on models that experience a DSF or begin to burn carbon.

Our 4.0, 4.5 and 5.0 $M_\odot$ models each experience a weak hydrogen-burning flash just after the thick shell helium burning phase with peak $L_H \sim 2\times10^6$ $L_\odot$. After 12 weak helium shell flashes during which the energy generation rate from helium burning reactions, $L_{He}$, at the peak of the flash does not exceed $6.3\times10^5$ $L_\odot$, the 4.0 $M_\odot$ model has a DSF, with a peak $L_H \sim 10^8$ $L_\odot$, which leads to dredge-up of CNO elements, mainly in the form of $^{12}C$. After the DSF, the peak value of $L_{He}$ during a flash grows to exceed $10^8$ $L_\odot$. Further dredge-up of $^{12}C$ occurs with each shell flash, and when we terminate our calculation after 79 flashes, the surface abundances are $X = 0.605$, $Y = 0.386$, $X_C = 7.3\times10^{-3}$, $X_N = 8.2\times10^{-4}$, $X_O = 3.3\times10^{-4}$, and $X_{Mg} = 2.5\times10^{-4}$. The $^{12}C/^{13}C$ number ratio is 42. The 4.5 $M_\odot$ model also experiences a DSF after a few weak He shell flashes. The 5.0 $M_\odot$ model does not experience a DSF. The first 100 He shell flashes are relatively weak with the peak value of $L_{He}$ during a flash not exceeding $10^5$ $L_\odot$. None of our 5.5, 6.0, 6.5 and 7.0 $M_\odot$ models has a DSF. The 5.5, 6.0 and 6.5 $M_\odot$ models enter the TP-AGB phase without there being a hydrogen flash after the thick shell helium burning phase. After the 11$^{th}$ helium shell flash in the 6.5 $M_\odot$ model, the bottom of the surface convection zone moves inward almost to the base of the helium layer which leads to a hydrogen flash and consequent dredge-up of CNO elements. Because the temperature at the bottom of the convection zone exceeds $10^8$ K, hot bottom burning (e.g. Lattanzio & Wood 2004) occurs and the dredge-up nuclei are mainly $^{14}N$. In agreement with the findings of Cassisi & Castellani (1993), off-center carbon burning begins in our 7.0 $M_\odot$ model before thermal pulses occur. After a carbon-depleted core of mass $\sim 0.9$ $M_\odot$ has formed, the star enters a super asymptotic giant branch (SAGB) phase (García-Berro & Iben 1994) during which the evolution of the thermal pulses is similar to that of the 6.5 $M_\odot$ model.

We find that the 4.0, 6.5 and 7.0 $M_\odot$ models, which all experience strong helium shell flashes, are all possibly net producers of $^7Li$. For example, if the $^7Li$ abundance remains the same as at the end of the last flash in our calculation, we estimate that our 4.0 $M_\odot$ model will eject $\sim 3\times10^{-8}$ $M_\odot$ of $^7Li$, which is more than the $\sim 1\times10^{-8}$ $M_\odot$ of $^7Li$ locked up in forming the star.

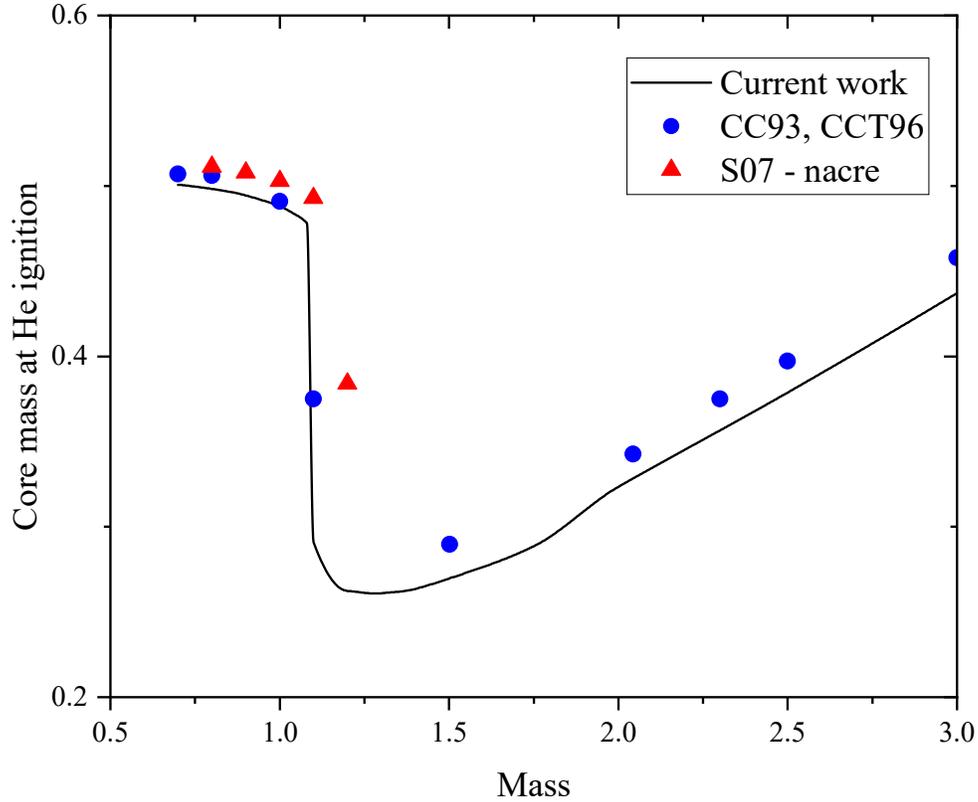

Figure 9. Comparison of the core mass at the onset of core helium burning from the current work with results from Cassisi & Castellani (1993), Cassisi et al. (1996) and Suda et al. (2007). The masses are in units of $M_\odot$.

## 3.5 Comparisons with previous work

Of particular interest for population III stars is the value of the initial mass dividing DCF and DSF behaviour. To refine this critical mass, we have evolved additional models of mass 1.05 and 1.08 $M_\odot$ up to the start of the DCF. In figure 9, we compare the mass of the hydrogen-exhausted core at the onset of core helium burning with results from Cassisi & Castellani (1993), Cassisi et al. (1996) and Suda et al. (2007). Despite a number of differences in the construction of the models, such as the initial He abundance, treatment of mass loss, the equation of state, etc., we see that there is good agreement between the various calculations. The transition from DCF to DSF behaviour occurs at the sharp drop in core mass which for our models lies between 1.08 and 1.10 $M_\odot$.

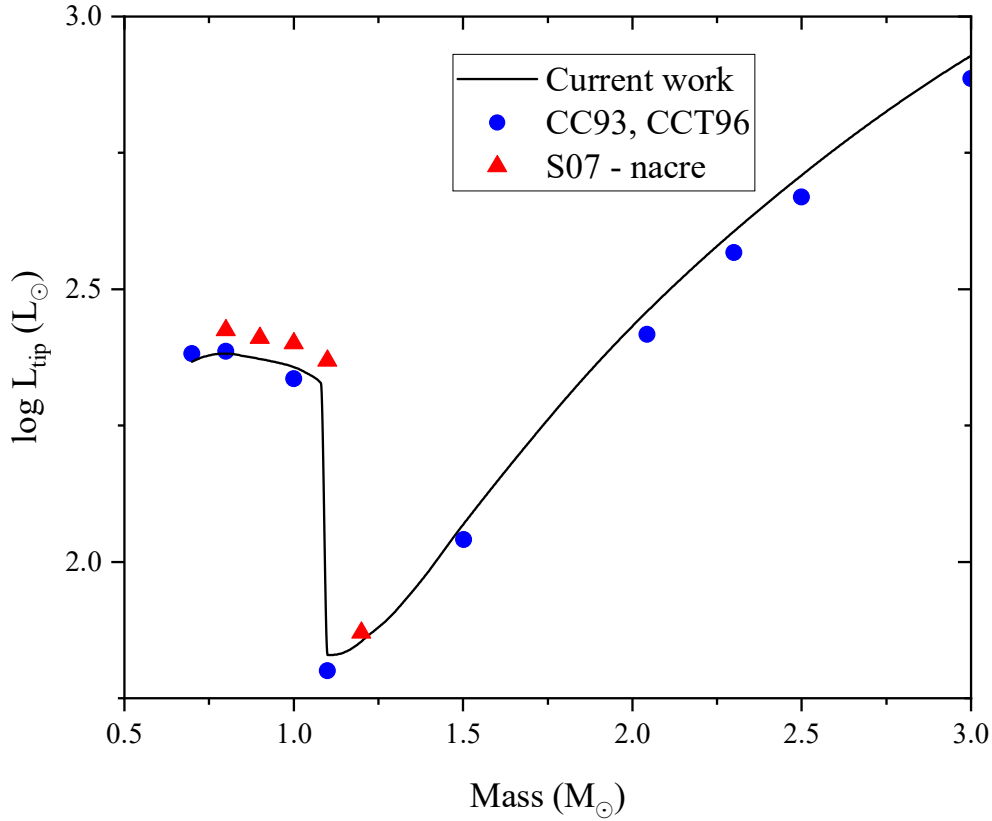

Figure 10. Comparison of the luminosity at the onset of core helium burning from the current work with results from Cassisi & Castellani (1993), Cassisi et al. (1996) and Suda et al. (2007).

We also compare in figure 10, the stellar luminosity at the 'RGB tip', which we take to be the red most point on the evolutionary track after the ZAMS but before the peak of the core flash for models that experience a DCF or the onset of quiescent helium core burning for those that do not have a DCF. Again, there is good agreement between the various models.

Cassisi et al. (1996) and Schlattl et al. (2001) have considered whether a significant number of RR Lyrae stars could be produced by an initially metal-deficient population of low mass progenitors. In the absence of a DCF, Cassisi et al. found that the ZAHB is always hotter than the RR Lyrae instability strip, and no ZAHB pulsators would be expected. According to Schlattl et al., inspection of the evolutionary paths in the H-R diagram shows that the possibility to form a significant number of variables from the more evolved stars is also negligible. Schlattl et al. find that even when for the DCF is taken into account, initially metal-free stars produce a negligible number of RR Lyrae variables, because now the ZAHB is too red when compared to the instability strip, and in post-ZAHB evolution the instability strip is crossed in a very short time (~1-2 Myr). They conclude that 'The only possibility for obtaining ZAHB pulsators is a quite strong but improbable mass loss rate (Reimers $\eta > 0.4$), which 'forces' the ZAHB location to lie within the instability strip.'

Inspection of figures 1 to 5 shows that the core helium burning phases for our models of mass 0.80, 0.85 and 0.90 $M_\odot$ are too blue for an RR Lyrae instability strip but, due to total mass loss rates corresponding to an effective Reimers $\eta > 0.4$, one could be possible for our models of mass 0.95 or 1.00 $M_\odot$. Interpolation in our models indicates that the ZAHB would lie in the instability strip for a narrow range of mass centred on 0.94 $M_\odot$ with width ~0.005 $M_\odot$. This would require the population to have age between 7.92 and 8.05 Gyr. The time spent crossing the instability strip in the post-ZAHB phase is 2.5 Myr and 3.7 Myr for our 0.95 or 1.00 $M_\odot$ models, respectively. This is longer than found by Schlattl et al. (2001) and hence the possibility of finding a Pop. III RR Lyrae star is increased by a factor of about 2. However, an estimate of the number of such stars would require knowledge of the initial mass function which is not well known.

## 4 Pre-WD surface abundances and white dwarf properties

In this section we provide our pre-WD surface abundances, and examine some differences between our population III WD's and earlier models with higher Z, including the initial – final mass relation (IFMR), and the final hydrogen and helium masses. Recently, Cummings et al. (2018) combined spectroscopic mass determinations of Sirius B and 79 cluster white dwarfs with application of PARSEC (Bressan et al. 2012) and MIST (Choi et al. 2016; Dotter 2016) isochrones to the clusters to determine an IFMR. The WDs in their study range from ~0.5 $M_\odot$ to ~1.3 $M_\odot$. Our initial - final mass relation (IFMR) is shown in figure 11, together with the semi-empirical IFMR from Cummings et al. (2018) that uses a fit to the empirical data based on solar abundance MIST isochrones (Choi et al. 2016; Dotter 2016), and the IFMR from Lawlor & MacDonald (2006) for their $Z = 0.02$ models. In general, the higher initial abundance $Z = 0.02$ models lead to lower mass white dwarf remnants, except notably for our models with $M_i \leq 0.90$ $M_\odot$, which have an extended hydrogen burning phase initiated by proton mixing during the helium core flash. We also see that the pop. III IFMR is not quite monotonic with a local maximum near $M_i = 1.0$ $M_\odot$ and a local minimum near $M_i = 2.0$ $M_\odot$.

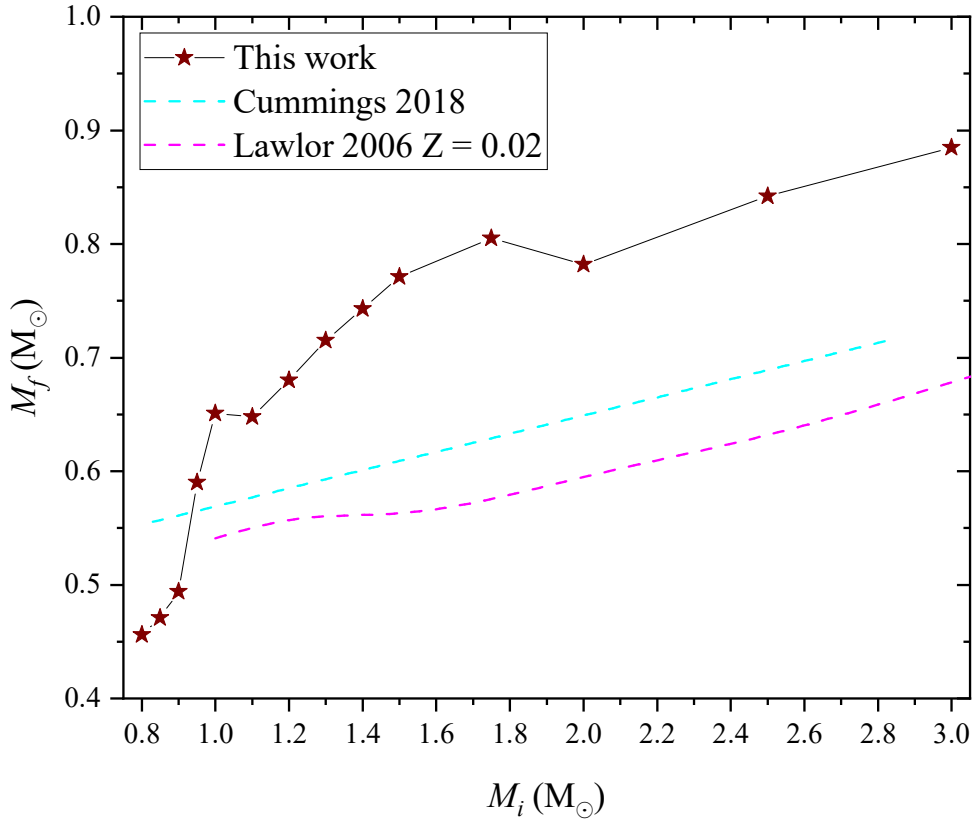

Figure 11. We compare our Population III initial-final white dwarf mass relation to the semi-empirical relation of Cummings et al. (2018) and the theoretical relation for the Z = 0.02 models of Lawlor & MacDonald (2006).

In Table 5, we give the surface composition parameters for our models just before entry onto the final white dwarf cooling track. The initial mass and final mass of each model are given in columns 1 and 2. The remaining columns give the mass fractions of H, He, $^{12}$C, $^{14}$N, $^{16}$O, and heavy elements, respectively. Of particular note are the 0.85 and 1.5 M$_\odot$ models which have large pre-white dwarf stage heavy element abundances. The mechanisms leading to the large Z values are different in the two cases. As described in more detail in section 3.2, the 0.85 M$_\odot$ model experiences significant mass loss during the post-DCF red giant phase and the secondary helium core flashes, so that when it settles on the horizontal branch it has an envelope mass of ~10$^{-2}$ M$_\odot$. Further mass loss uncovers the heavy element rich layers. The high Z value for the 1.5 M$_\odot$ model is due to a very late thermal pulse (VLTP).

Table 5. Surface composition parameters just before the white dwarf stage

| $M_i$ (M$_\odot$) | $M_f$ (M$_\odot$) | X | Y | $^{12}C$ | $^{14}N$ | $^{16}O$ | $Z_f$ |
|---|---|---|---|---|---|---|---|
| 0.80 | 0.456 | 0.479 | 0.506 | 4.42 10$^{-3}$ | 7.24 10$^{-3}$ | 1.24 10$^{-3}$ | 0.0144 |
| 0.85 | 0.471 | 0.024 | 0.931 | 2.79 10$^{-2}$ | 6.85 10$^{-3}$ | 3.68 10$^{-3}$ | 0.0450 |
| 0.90 | 0.494 | 0.523 | 0.456 | 3.18 10$^{-3}$ | 6.01 10$^{-3}$ | 9.13 10$^{-4}$ | 0.0112 |
| 0.95 | 0.590 | 0.546 | 0.443 | 1.73 10$^{-5}$ | 9.89 10$^{-3}$ | 8.04 10$^{-4}$ | 0.0108 |
| 1.00 | 0.651 | 0.559 | 0.432 | 1.63 10$^{-3}$ | 6.63 10$^{-3}$ | 7.11 10$^{-4}$ | 0.0096 |
| 1.10[a] | 0.648 | 0.698 | 0.300 | 6.91 10$^{-5}$ | 7.47 10$^{-4}$ | 3.39 10$^{-5}$ | 0.0009 |
| 1.10[b] | 0.648 | 0.541 | 0.428 | 2.76 10$^{-2}$ | 7.28 10$^{-4}$ | 1.82 10$^{-3}$ | 0.0308 |
| 1.20 | 0.680 | 0.688 | 0.311 | 9.77 10$^{-5}$ | 7.93 10$^{-4}$ | 3.81 10$^{-5}$ | 0.0010 |
| 1.30 | 0.715 | 0.684 | 0.315 | 9.13 10$^{-5}$ | 7.40 10$^{-4}$ | 3.18 10$^{-5}$ | 0.0009 |
| 1.40 | 0.743 | 0.677 | 0.321 | 3.17 10$^{-4}$ | 7.08 10$^{-4}$ | 4.22 10$^{-5}$ | 0.0011 |
| 1.50[c] | 0.771 | 0.673 | 0.326 | 5.37 10$^{-4}$ | 6.83 10$^{-4}$ | 4.73 10$^{-5}$ | 0.0013 |
| 1.50[d] | 0.771 | 0.065 | 0.690 | 1.50 10$^{-1}$ | 3.00 10$^{-2}$ | 3.49 10$^{-2}$ | 0.2441 |
| 1.75 | 0.805 | 0.669 | 0.325 | 4.46 10$^{-3}$ | 3.88 10$^{-4}$ | 2.10 10$^{-4}$ | 0.0052 |
| 2.00 | 0.782 | 0.676 | 0.321 | 2.26 10$^{-3}$ | 2.13 10$^{-4}$ | 8.51 10$^{-5}$ | 0.0026 |
| 2.50 | 0.842 | 0.643 | 0.352 | 3.69 10$^{-3}$ | 2.30 10$^{-4}$ | 1.40 10$^{-4}$ | 0.0041 |
| 3.00 | 0.885 | 0.619 | 0.375 | 5.88 10$^{-3}$ | 2.56 10$^{-4}$ | 2.51 10$^{-4}$ | 0.0066 |

[a] Before LTP, [b] After LTP, [c] Before VLTP, [d] After VLTP

Asteroseismology has proven to be a powerful technique for probing in the interior structure of stars (Kurtz 2022). Application to WDs allows determination of stellar mass, hydrogen and helium layer masses, rotation rates, and, in some cases, the rate of cooling and upper limits on magnetic field strength (Fontaine & Brassard 2008; Winget & Kepler 2008). In table 6, we give $M_H$ and $M_{He}$, the masses of hydrogen and helium in our white dwarf models, respectively, for comparison with potential asteroseismology analyzes. Our models of initial mass 0.85 M$_\odot$ and 1.5 M$_\odot$ have white dwarf hydrogen masses an order of magnitude lower than other comparable models, for reasons given above.

Table 6. Masses of hydrogen and helium in our white dwarf models

| $M_i$ (M$_\odot$) | $M_f$ (M$_\odot$) | $M_H$ (M$_\odot$) | $M_{He}$ (M$_\odot$) |
|---|---|---|---|
| 0.80 | 0.456 | 1.69 10$^{-4}$ | 0.453 |
| 0.85 | 0.471 | 5.82 10$^{-6}$ | 5.39 10$^{-2}$ |
| 0.90 | 0.494 | 1.49 10$^{-4}$ | 4.58 10$^{-2}$ |
| 0.95 | 0.590 | 8.12 10$^{-5}$ | 2.38 10$^{-2}$ |
| 1.00 | 0.651 | 5.45 10$^{-5}$ | 1.59 10$^{-2}$ |
| 1.10 | 0.648 | 3.15 10$^{-5}$ | 6.29 10$^{-3}$ |
| 1.20 | 0.680 | 6.60 10$^{-5}$ | 8.35 10$^{-3}$ |
| 1.30 | 0.715 | 5.19 10$^{-5}$ | 5.83 10$^{-3}$ |
| 1.40 | 0.743 | 4.09 10$^{-5}$ | 3.68 10$^{-3}$ |
| 1.50 | 0.771 | 5.01 10$^{-7}$ | 2.06 10$^{-3}$ |
| 1.75 | 0.805 | 1.82 10$^{-5}$ | 3.48 10$^{-3}$ |
| 2.00 | 0.782 | 2.58 10$^{-5}$ | 3.32 10$^{-3}$ |
| 2.50 | 0.842 | 1.44 10$^{-5}$ | 2.05 10$^{-3}$ |
| 3.00 | 0.885 | 9.17 10$^{-6}$ | 1.21 10$^{-3}$ |

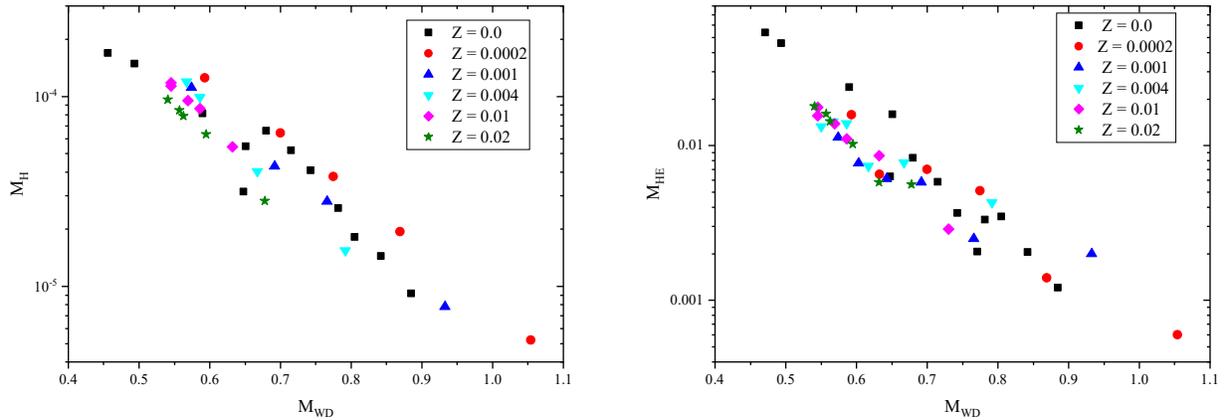

Figure 12. Comparison of white dwarf hydrogen and helium masses for Z = 0.0 models (current work) with those of Lawlor & MacDonald (2006) for a range of initial heavy element abundances.

In figure 12, we compare the WD hydrogen and helium masses with those found by Lawlor & MacDonald (2006) for models in a range of heavy element abundances from Z = 0.0002 to 0.02. We see that the hydrogen mass for pop III models is not distinguishable from the hydrogen mass for higher metallicity models. However, for white dwarfs less massive than ~0.65 M$_\odot$, the helium mass for pop III models is higher than for the higher metallicity models, which might allow asteroseismology to distinguish the white dwarf descendants of single pop III stars of initial mass less than ~1.0 M$_\odot$ from similar mass white dwarf descendants of more metal-rich progenitors.

## 5 SUMMARY AND DISCUSSION

We have calculated stellar evolution models of stars with primordial initial composition for a range of low to intermediate mass. For $M \leq 3.0$ M$_\odot$, the evolutionary tracks are followed to the white dwarf stage. We have also investigated the evolution of higher mass models to the TP-AGB phase. We have included the effects of mass loss by adding to the Reimers' mass loss rate a prescription for mass loss from red giant stars that have dusty envelopes.

We find that single Pop. III stars of initial mass greater than 0.81 M$_\odot$ evolve to the white dwarf stage in 13.5 Gyr or less. As in some earlier studies (Fujimoto et al. 1990; Hollowell et al. 1990; Fujimoto et al. 2000; Schlattl et al. 2001; Suda et al. 2007; Campbell & Lattanzio 2008), we find that models of $M \leq 1.0$ M$_\odot$, experience a double core flash in which the convection zone driven by the helium core flash reaches into the bottom layers of the hydrogen-rich envelope. However, unlike the earlier studies, we also find that after the DCF our models evolve further to the red due to an increase of the opacity in the envelope that results from dredge up of CNO elements. Models with initial mass $\leq 0.95$ M$_\odot$ experience a phase near the tip of the giant branch in which mass loss is mostly due to a dusty wind, which combined with an extended period of hydrogen burning that follows the DCF can lead to significant total mass loss so that the model becomes a relatively low mass white dwarf. Due to mass loss, we find that none of our models that have a DCF enters the TP-AGB phase. Except for the 0.85 M$_\odot$ model, the surface composition for evolutionary stages after the DCF is essentially determined by the convective dredge-up that occurs shortly after the DCF. The dominant surface element is $^{14}$N, with a $^{14}$N/$^{12}$C ratio between ~2 to ~4. Mass loss during the secondary helium core flashes leads to removal of most of the envelope of our 0.85 M$_\odot$ model, exposing material in which the 3α reaction has produced significant amounts of $^{12}$C.

Models with mass in the range $1.1 \leq M \leq 4.5$ M$_\odot$ all have a DSF, whereas models of mass $\geq 5.0$ M$_\odot$ do not. The main isotope dredged up immediately after the DSF is $^{14}$N but for models in the mass range $1.75 \leq M \leq 4.5$ M$_\odot$ further dredge up episodes during the TP-AGB phase lead to $^{12}$C becoming the dominant CNO isotope at the stellar surface. Off-center carbon-burning occurs in our 7.0 M$_\odot$ model before the TP-AGB phase. After a carbon-depleted core of mass ~0.9 M$_\odot$ has developed, this model enters a Super-AGB phase.

Stars of very low heavy element abundance are of interest because they provide information on the formation and evolution of the Galaxy, and also set a limit on its age and the age of the Universe. As yet no Pop. III stars have been definitively identified in our Galaxy or elsewhere in the Universe. Vanzella et al. (2020) discovered a strongly lensed star cluster which they find is consistent with being a complex of population III stars at $z = 6.6$. Support for this idea is provided by the theoretical prediction (Trenti, Stiavelli, & Shull 2009) that pockets of zero-metallicity gas, from which population III stars can form, may still exist at $z \approx 6$, particularly near the outskirts of galaxies. Welch et al. (2022) have considered that the most distance star, WHL0137-LS (Earendel) at redshift 6.2, that they discovered with the Hubble Space telescope could be a Pop. III star.

In general, the stars in our Galaxy with the lowest known heavy element abundances ([Fe/H] < -5) are carbon-enhanced metal poor (CEMP) stars. Can the carbon-enhancement in any of these stars be produced by dredge-up following a DCF in a pop. III star? Ideally the iron abundance should be zero or very low in the case that accretion from the ISM has slowly occurred over the lifetime of the star (Shigeyama, Tsujimoto & Yoshii 2003). From the evolutionary tracks shown in figures 1 – 5, we see that the core flash occurs at $T_{eff} \sim 4900$ K, and hence any CEMP star cooler than this limit is a possible pop. III candidate. Also, CEMP stars created by the DCF in population III should have strong nitrogen enhancement, and no s-process elements.

The lowest metallicity star known is SMSS J031300.36-670839.3 (Keller et al. 2014) in which no iron lines have been detected and has an upper limit on its metallicity of [Fe/H] < -7.1. Detected elements have abundances [C/H] = -2.6, [Mg/H] = -3.8 and [Ca/H] = -7.0 with an upper limit on the nitrogen abundance of [N/H] = -3.9. The abundance of neutral Li is A(Li I) = 0.7. The best estimates for photospheric parameters are $T_{eff}$ = 5,125 K, and log $g$ = 2.3, which have been confirmed by Sitnova et al. (2019). These atmospheric parameters place SMSS J031300.36-670839.3 squarely on the RGB before the core helium flash, and hence it is unlikely to be a Pop. III star. The non-detection of nitrogen also argues against a population III star progenitor.

The next lowest [Fe/H] abundance is found for SMSS J160540.18−144323.1 (Nordlander et al. 2019), which has [Fe/H] = −6.2 ± 0.2, [C/H] = -2.3 ± 0.2, [Ca/Fe] = 0.4 ± 0.2, [Mg/Fe] = 0.6 ± 0.2, [Ti/Fe] = 0.8 ± 0.2, and no significant s- or r-process enrichment. The upper limit on the nitrogen abundance is [N/H] < -3.0. Nordlander et al. determine $T_{eff}$ = 4850 ± 100 K, log $g$ = 2.0 ± 0.2. Although $T_{eff}$ is low enough that this star could be a post-DCF population III star, the lack of nitrogen is problematical for this scenario.

Other CEMP stars with low [Fe/H] abundances are HE 1327−2326 which has [Fe/H] = −5.7 (Frebel et al. 2005; Aoki et al. 2006), HE 0107−5240 for which [Fe/H] = −5.4 (Christlieb et al. 2002, 2004) and SD 1313−0019 at [Fe/H] = −5.0 (Allende Prieto et al. 2015; Frebel et al. 2015). For HE 1327−2326, Aoki et al. estimate $T_{eff}$ = 6180 K, which is too hot for carbon enrichment in a DCF to have occurred. Christlieb et al. (2004) determined that HE 0107−5240 has $T_{eff}$ = 5100 ± 150 K, [C/H] = -1.28 based analysis of a $C_2$ band, or [C/H] = -1.58 from CH lines, [N/H] = -3.00 or -2.71 depending on the C abundance, a $^{12}C/^{13}C$ ratio of ~60. Na, Mg, Ca, Ti, Fe, and Ni are also present in detectable amounts. With just an upper limit of [Ba/Fe] < 0.82 on the barium abundance it is not clear whether s-process elements are present. Bessell, Christlieb & Gustafsson (2004) by using a plane-parallel 1D LTE model atmosphere to analyze OH limes in ultraviolet spectra determined [O/Fe] = 2.3, and estimate systematic errors due to three-dimensional effects would reduce the oxygen abundance by the order of 0.3–0.4 dex. In terms of mass fractions, the abundances of CNO elements are approximately $X_C$ = 2 $10^{-4}$, $X_N$ = 2 $10^{-6}$, $X_O$ = 2 $10^{-5}$. As pointed out by Limongi, Chieffi & Bonifacio (2003), this abundance pattern is inconsistent with a self-pollution scenario due to a DCF. Suda et al. (2004) have proposed that the abundance patterns observed in HE 0107−5240 are due to the Fe and other heavy elements being accreted by a pop. III star from a primordial cloud that has been seeded by supernovae followed by accretion of the lighter elements due to wind mass transfer from a more massive companion in which the carbon is produced by dredge-up episodes during the TP-AGB. In this scenario, which has similarities to the formation mechanism proposed for CH and Ba stars (McClure, Fletcher & Nemec 1980; Boffin & Jorissen 1988), the companion star evolves to a white dwarf in an orbit of semi-major axis ~34 AU and period of ~150 yr. However, Bessell at al. argue that it is unlikely that the large CNO overabundances in the atmosphere of HE 0107-5240 have been produced in such a way, because they note that typical orbital periods of CEMP stars known to be binaries are 1.5 – 8.5 yr, and state that in shorter period systems, mass transfer would occur before the initially more massive companion reaches the AGB phase, and in longer period systems, mass transfer does not occur at all. However, Jorissen et al. (2016) find an orbital period of 54 ± 7 yr for the CH-star HD 26, and do not rule out that the period might even be longer. Although Frebel et al. (2015) find SD 1313−0019 does have a significant nitrogen abundance with [N/C] = 0.5 and no evidence for s-process elements, they determine $T_{eff}$ = 5200 ± 150 K, which is likely too high for a DCF to have occurred. So, it seems, with the possible exception of HE 0107-5240, that none of these stars with [Fe/H] < -5 is primordial.

Age determinations of the oldest stars in our Galaxy indicate that star formation in its building blocks started very shortly after the big bang (e.g. Cowan et al. 2002; Bond et al. 2013). Assuming that low mass population III stars did form early, those of mass greater than ~0.81 $M_\odot$ will have evolved to the white dwarf stage.

White dwarf stars have often been used for cosmochronometry measurements. Field white dwarfs have been used to determine the age of the various components of the Milky Way. For example, Kilic et al. (2017) used the luminosity function of a sample of local white dwarfs (within 40 pc) to determine ages of 6.8–7.0 Gyr for the thin disk and 8.7 ± 0.1 Gyr for the thick disk, and an age estimate of $12.5^{+1.4}_{-3.4}$ Gyr for the inner halo. More recently, Fantin et al. (2021) have analysed their data for 18 halo white dwarfs to determine their masses and ages. Based on their determination of a mean age of $9.03^{+2.13}_{-2.03}$ Gyr and a dispersion of $4.21^{+2.33}_{-1.58}$ Gyr, Fantin et al. suggest that there was an extended star formation history within the local halo population.

Underpinning the Kilic et al. age estimates are the initial–final mass relation from Kalirai et al. (2008) and main sequence lifetimes from the analytic formulae for stellar evolution properties as a function of mass and metallicity of Hurley, Pols & Tout (2000) found by fitting to the detailed constant mass evolutionary tracks calculated by Pols et al. (1998) for $Z$ in the range $10^{-4}$ to 0.03. Because of the large differences in behaviour of Pop. III stellar models and models with $Z = 10^{-4}$ and higher, it would be inappropriate to use the analytic formulae to determine the age of any white dwarf stars in the Milk Way that might have had Pop. III progenitors.

Our models provide constraints on the ages of halo white dwarfs, and are particularly useful for stars that have recently entered the white dwarf cooling phase. Fantin et al. (2021) find that the lowest mass DA white dwarf in their sample, J0823+3111, has an age of 14.6 ± 2.2 Gyr based on MIST models and 14.8 ± 2.3 Gyr when PARSEC models are used. In both cases, the Cummings et al. (2018) IFMR was used to determine the progenitor mass. Although these age estimates are consistent with the age of the Universe within the error bars, the fact that the mean ages are higher than the age of the Universe suggest that progenitor might have lower metallicity than the models used in determination of the age and progenitor mass. If we use our population III models, we find age of 11.2 ± 2.2 Gyr, which is in better agreement with the age of the Universe.

ACKNOWLEDGEMENTS

We thank the referee for their comments and suggestions for improving the manuscript. We gratefully acknowledge discussion with Volker Bromm in regard to the possibility of the formation of low mass population III stars. This work was partially supported by the NASA Delaware Space Grant Program.

DATA AVAILABILITY

No new data were generated or analyzed in support of this research.